\def\year{2022}\relax
\documentclass[letterpaper]{article} % DO NOT CHANGE THIS
\usepackage{aaai22}  % DO NOT CHANGE THIS
\usepackage{times}  % DO NOT CHANGE THIS
\usepackage{helvet}  % DO NOT CHANGE THIS
\usepackage{courier}  % DO NOT CHANGE THIS
\usepackage[hyphens]{url}  % DO NOT CHANGE THIS
\usepackage{graphicx} % DO NOT CHANGE THIS
\usepackage{natbib}  % DO NOT CHANGE THIS AND DO NOT ADD ANY OPTIONS TO IT
\usepackage{caption} % DO NOT CHANGE THIS AND DO NOT ADD ANY OPTIONS TO IT
\DeclareCaptionStyle{ruled}{labelfont=normalfont,labelsep=colon,strut=off} % DO NOT CHANGE THIS
\usepackage{algorithm}
\usepackage{algorithmic}
\usepackage{subcaption}
\usepackage{amsmath}
\usepackage{amssymb}

\usepackage{newfloat}
\usepackage{listings}
\floatstyle{ruled}
\newfloat{listing}{tb}{lst}{}
\floatname{listing}{Listing}
\title{Automated Vision-Based Wellness Analysis for Elderly Care Centers}
\author{
    Xijie Huang, Jeffry Wicaksana, Shichao Li, Kwang-Ting Cheng\\
}
\affiliations{
    Hong Kong University of Science and Technology\\

    \{xhuangbs, jwicaksana, slicd\}@connect.ust.hk, timcheng@ust.hk
}

\begin{document}

\maketitle

\begin{abstract}
The growth in the aging population require caregivers to improve both efficiency and quality of healthcare. 
In this study, we develop an automatic, vision-based system for monitoring and analyzing the physical and mental well-being of senior citizens. 
Through collaboration with Haven of Hope Christian Service, we collect video recording data in the care center with surveillance camera. We then process and extract personalized facial, activity, and interaction features from the video data using deep neural networks.
This integrated health information systems can assist caregivers to gain better insights into the seniors they are taking care of. 
These insights, including wellness metrics and long-term health pattern of senior citizens, can help caregivers update their caregiving strategies.
We report findings of our analysis and evaluate the system quantitatively. 
We also summarize technical challenges and additional functionalities and technologies needed for offering a comprehensive system.

%The rapid growth in the aging population yet limited manpower to provide essential elderly care calls for new tools to assist caregivers in improving both efficiency and quality of care in elderly care centers. In this study, we develop an automatic, vision-based system for analyzing and monitoring the physical and mental well-being of senior citizens and for assisting staff to gain better insights into the seniors they are taking care of. Through collaboration with Haven of Hope Christian Service, we developed a real-time video content analysis system using which we conducted case studies to analyze the activities and social interactions of senior citizens with dementia in their daycare centers. In particular, the activities that the current system targets to identify include napping, sitting, yawning, watching TV, eating, exercising, and talking. Daily analysis of individual senior’s activity patterns and respective duration, and their intensities of the guided exercises becomes feasible. Such analysis can reveal senior’s daily routines, as well as the long-term patterns and trends. We report evaluation results and findings of our analysis based on comprehensive video data captured at the elderly care centers. We also summarize challenges and additional functionalities and technologies needed for offering a comprehensive system.

\end{abstract}

\section{Introduction}
The need to rethink the care quality for elderly citizens is getting more important as the growth of the older population outpaces the growth of available caregivers. During the COVID-19 pandemic, there have been several outbreaks in the elderly care centers ~\cite{heras2021covid, gardner2020coronavirus} because caregivers need to serve multiple care centers. To address the manpower issue, we aim to improve the efficiency of each caregiver by providing them a vision-based tool to automatically provide assistive insights about each senior citizen.

%With the advance of artificial intelligence technologies, deep learning achieves significant performance enabling many vision applications. 

The representational power brought by the development of deep learning approaches and increased amount of annotation has revolutionized the way of processing visual data. Such visual understanding models lead to new opportunities for human-centric healthcare. For example, human activity classification from video data can be extended to classify clinically relevant activities and monitor the occurrence of abnormal events such as falls~\cite{rougier2011robust, yu2012posture, mastorakis2014fall, zhang2015survey}.

\begin{figure}[t]
	\begin{center}
		\includegraphics[width=0.5\textwidth]{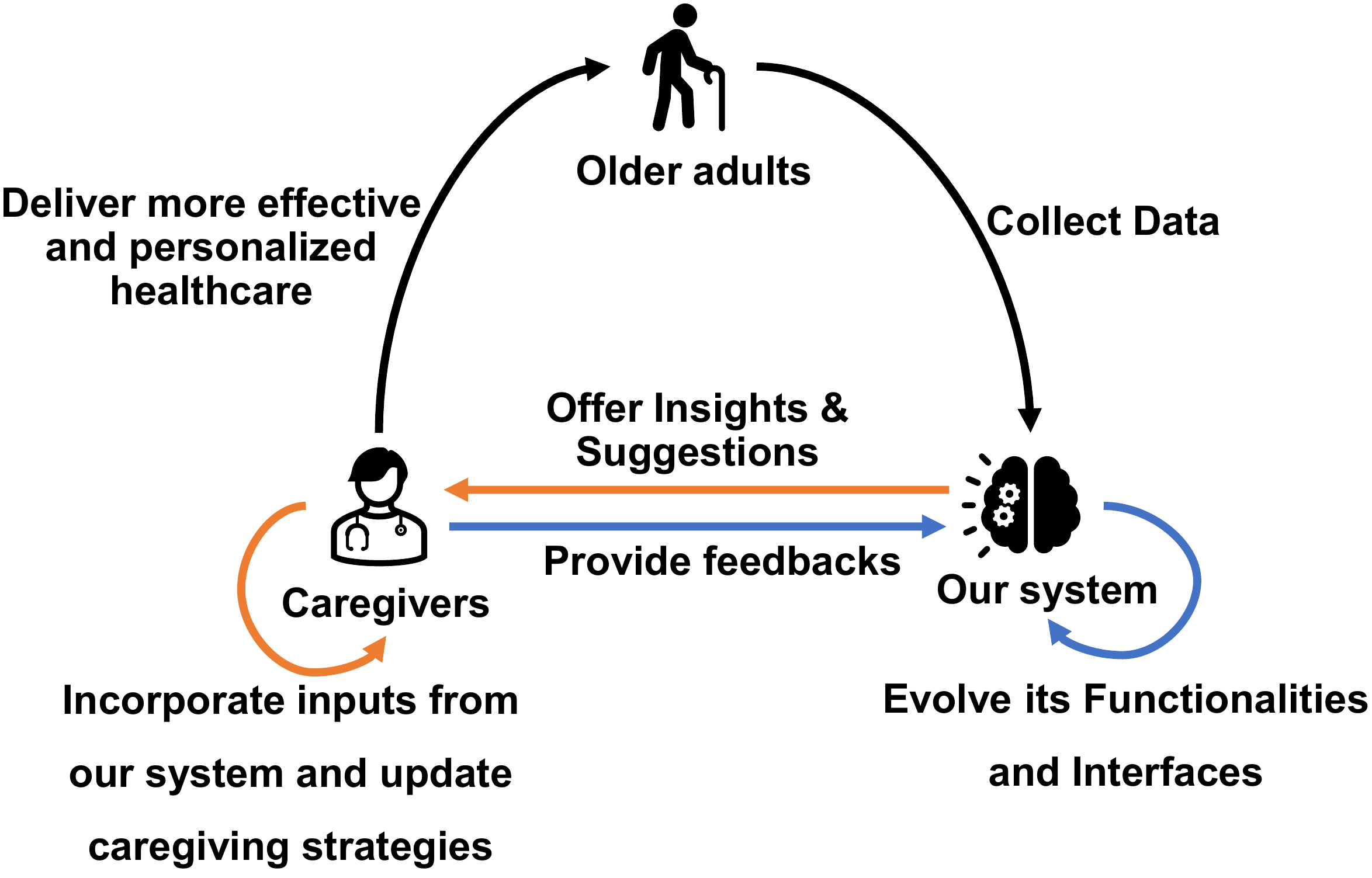}
	\end{center}
	\caption{\textbf{Value proposition Diagram}: The goal of our system is to assist caregivers to deliver more effective and personalized care to older adults. We achieve this goal by designing a system that derives insights from raw data of older adults and offers meaningful suggestions to caregivers.}
	\label{Figure:value}
\end{figure}

To improve elderly healthcare quality through computer vision-based solution, we partner with the Haven of Hope Christian Service elderly care centers in Hong Kong to develop such a system. The value proposition of our system is demonstrated in Fig.~\ref{Figure:value}. Cameras were installed to capture the daily activities of senior citizens which are automatically analyzed with the goal of providing meaningful insights relevant to senior citizens' wellness and their trends. Results are then provided to caregivers to adjust their care strategy regarding who, when, and where they should pay greater attention to. We collected about one month of video data from one daycare center of demented elder to drive the development and evaluation of our automated vision-based system.

\begin{figure*}[t]
	\begin{center}
		\includegraphics[width=\textwidth]{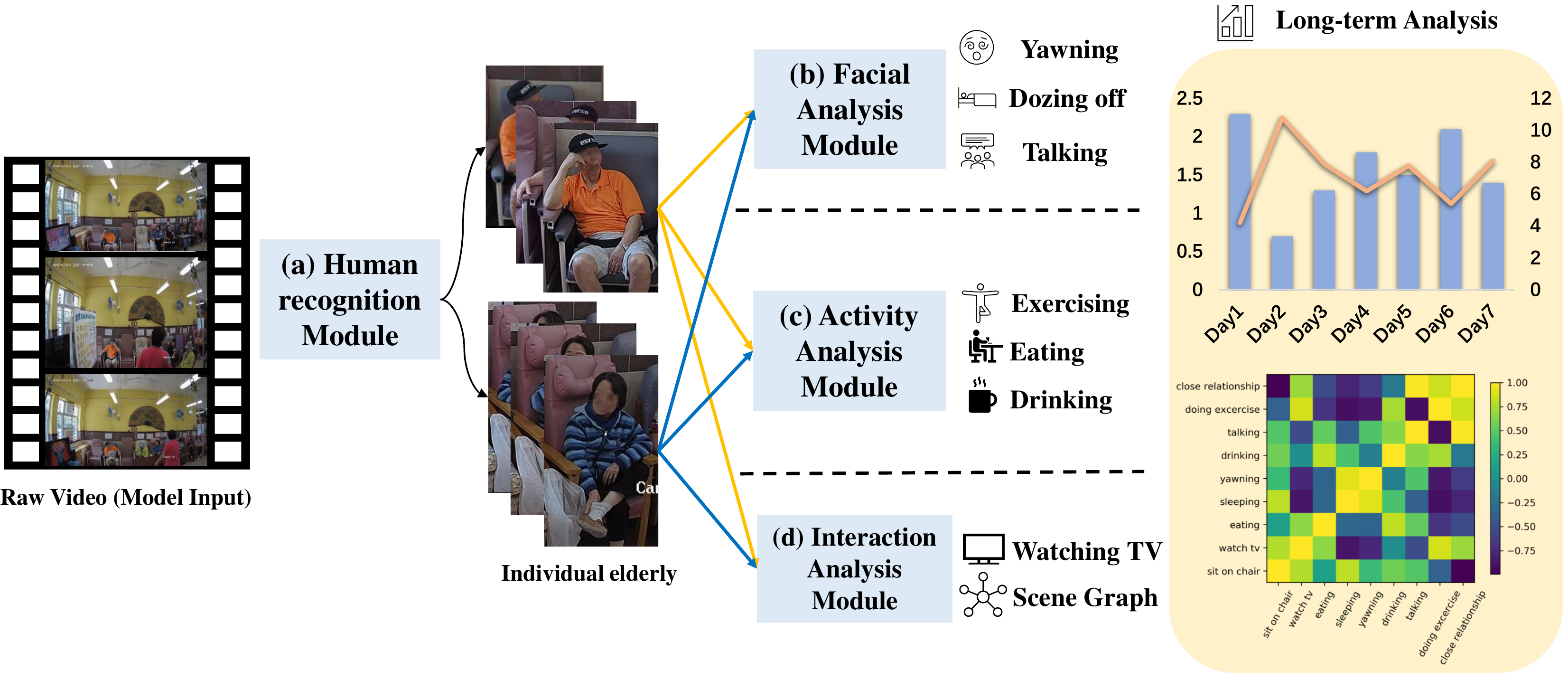}
	\end{center}
	\caption{Overview of our framework: a human recognition module detects individual elderly from the video clips, then feed the individual videos into three modules: facial analysis, activity detection, and scene understanding. The final target is to perform analysis, both daily and long-term, and provide immediate assistance for any detected anomaly.}
	%Interaction with others and environment
	\label{Figure:pipeline}
\end{figure*}

The overall framework of our system is illustrated in Fig.~\ref{Figure:pipeline}. After automatically identifying each individual, we analyze three major aspects of each individual senior citizen: the person's facial information, physical activities, and social interaction. Facial activities include but not limited to facial expression, yawning, blinking, and talking. The derived information with respect to these three components are then summarized and visualized to help caregivers better understand senior citizens’ daily behaviors and patterns over time. Additionally, detected anomaly that requires immediate assistance will also be provided to the caregivers.

In the care center, daily guided physical exercises are designed for senior citizens to keep them physically active. In order to evaluate and improve the efficacy of daily exercises and better tailor them for different senior citizens, we automatically analyze the physical movements of each person during the exercises. We define a metric, exercise intensity score, to summarize how active each senior citizen is when performing the physical exercises. We track the activeness level of each individual over time and when there is a significant deviation from their standard activeness level, caregivers can be alerted automatically. 

%The visualization of our current system consists of two components: activity temporal heatmaps and scene graphs. 

Our current system can produce two types of insightful visualization of the analysis results: the activity temporal heatmaps and the scene graphs. A temporal heatmap records various activities and their duration engaged by each person throughout a day. By tracking this information over a month, we gain significant insights to the behavior patterns of a senior citizen and are able to capture the moment or period when an individual is in distress. For instance, an individual naps, when sitting on a chair, much longer than his or her usual napping period, caregivers can be alerted. 

A scene graph focuses on capturing two aspects. First is the social interaction among individuals, including both senior citizens and caregivers. Second is the interaction between the elderly and inanimate objects such as television. Through our observation, senior citizens have their circle of friendships and preferred caregivers. 

% this para focus on why system can provide insights
Based on these visualization, caregivers can better update their caregiving strategy accordingly. Instead of monitoring the status of a single elderly, caregivers are able to monitor and compare the wellness of a group of elderly with our system. In the long term, caregivers will be notified when there is abnormal change of temporal health metric of individual and personalized care can be provided.

To quantitatively evaluate the detection and recognition accuracy of our system, we label some clips of video recording for various activities. We designed experiments to show that our system can achieve sufficient accuracy and efficiently detect various activities. In summary, this paper contributes as the following:
\begin{itemize}
    \item We report an automated vision-based wellness analysis system which can effectively and accurately perform facial analysis and activity analysis at the same time under the realistic elderly care scenarios.
    \item We propose visualization of the patterns via scene graphs and temporal activity heatmaps to provide an easy-to-understand summary for caregivers.
    \item We incorporate quantified social relationship as one important factor of mental wellness and apply data mining techniques to find the association of social activeness and other physical health indicators.
    \item We cooperate with Haven of Hope Christian Service elderly care center, collect and label a practical video dataset and evaluate the key component of our system quantitatively. 
  
\end{itemize}

\section{Related Work} %correspond to the contribution

Smart healthcare systems have drawn increased research attention since the flourish of artificial intelligence and smart devices. However, most of the existing healthcare systems require wearable devices to monitor the activity which is intrusive and hence not widely accepted. Computer vision based solutions bring the opportunity of a non-intrusive system based upon stationary cameras that allow passive detection of important activities. Most of the vision-based works focuses on fall detection~\cite{rougier2011robust, yu2012posture, mastorakis2014fall, zhang2015survey}. While these models are effective for identifying critical acute conditions, they do not detect and analyze daily behaviors and their long-term patterns, which are of great importance for understanding elderly citizens. 
Prior works focusing on the long-term health monitoring~\cite{parajuli2012senior, luo2018computer}, focuses on using depth~\cite{parajuli2012senior} and thermal~\cite{luo2018computer} sensors instead of vision sensors. To the best of our knowledge, an automated non-intrusive vision-based system that is capable of activity detection and long-term health monitoring has not been proposed prior to our work. 

An automated vision-based healthcare system calls for a comprehensive detection and analysis of individual condition and the surrounding circumstance, which includes facial analysis, action analysis, interaction analysis. We will introduce research related to this three topics in following subsections.

\subsection{Facial Analysis}

Recognizing people and analyzing their facial landmark and expression is important for a personalized healthcare system. 
A deep understanding of individual facial information may provide meaningful health-related insights.
For example, diagnosing Parkinson Disease via facial expression recognition~\cite{jin2020diagnosing, bandini2017analysis, alhussein2017monitoring} has been investigated and researchers have achieved satisfactory classification results. Moreover, monitoring neurological disorders~\cite{yolcu2017deep} and early Alzheimer's Disease (AD) diagnosis~\cite{bi2019early} based on facial features have also been studied. Targeting healthcare applications, Jin~\textit{et~al.} demonstrates the powerful representative capacity of facial feature via transfer learning from face recognition to facial diagnosis.

\subsection{Action Analysis}
The action analysis in our system involves the techniques of spatio-temporal action recognition in video and human-object interaction detection.
For spatio-temporal action recognition, Ji~\textit{et~al.}~\cite{ji20123d} first proposed 3D CNN for human action recognition which adopts consecutive frames of videos as inputs and extends the 2D convolutional filters to 3D via adding the temporal dimension. Karparthy~\textit{et~al.}~\cite{karpathy2014large} then verified and compared different fusion methods that utilize the temporal information in the videos. Based on VGGNet~\cite{simonyan2014very}, C3D~\cite{tran2015learning} was proposed containing 3D pooling layers and 3D convolutional filters together. 

Human object interaction (HOI) detection is adapted from a visual relationship detection problem. InteractNet~\cite{gkioxari2018detecting} is proposed to locate the interacted objects based on action-specific attention map estimation methods. Chao~\textit{et~al.}~\cite{chao2018learning} proposed a multi-stream model which fuses visual features and spatial locations. In~\cite{zhou2021cascaded}, Qi~\textit{et~al.} proposed a graph neural network-based method that iteratively updates states and classifies all possible pairs/edges representing the interaction. Gao~\textit{et~al.}~\cite{gao2018ican} proposed iCAN, containing an instance-centric attention module to enhance the HOI classification.

\subsection{Interaction Analysis}

Scene graph structure is widely adopted to represent the scene and we utilize scene graph generation (SGG) for the scene understanding. Xu~\textit{et~al.}~\cite{xu2017scene} proposed a message-passing model that iteratively refines the relation prediction. Zellers~\textit{et~al.}~\cite{zellers2018neural} introduced BiLSTM to encode visual context which can extract global visual features and improve relationship predictors. Chen~\textit{et~al.}~\cite{chen2019knowledge} proposed a knowledge-embedded routing network (KERN) that utilizes the statistical correlations to regularize semantic space and make prediction less ambiguous. Neural motifs~\cite{zellers2018neural} use GloVe~\cite{pennington2014glove} to embed word vectors to combine the semantic feature with the visual context.

\section{Proposed Wellness Analysis System}
In this section, we will show components of proposed system: facial analysis, activity analysis, and interaction analysis module. These three modules all target at extracting meaningful feature. We utilize visualization tool to demonstrate these feature as straightforward wellness metrics. 

\subsection{Facial Analysis} % use a what-how structure
To help caretakers to gain better understanding of the senior citizens they care for, we propose to detect their facial activities, including yawning, dozing off, talking,  blinking, etc. These facial activities can be detected by analyzing the facial landmarks dynamic between frames. 
We first briefly introduce how we extract the facial landmarks using a pre-trained deep neural network. Then we introduce how we utilize the facial landmarks to detect specific activities using a set of rules. 

% The facial feature is the most widely accepted biometric that can be applied to human recognition. 
For our application, the subjects of interest are senior citizens and caregivers. 
We built a human face library for all the subjects and apply the YOLOFace~\cite{li2020face} model to detect the faces. The model is built on YOLOv3~\cite{redmon2018yolov3} and trained on the WIDER FACE dataset~\cite{yang2016wider}. 
When a face is detected, we can pair each detected face with the corresponding detected human with maximization of the Intersection of Union (IoU). 
The next step for face feature extraction is detecting facial landmarks on the facial image obtained from the face detection model. 
We use 68 semantic landmarks defined in iBUG~\cite{sagonas2016300}. These landmarks are chosen as they include relevant facial features (eyes, nose, mouth, etc.) while being relatively light as well as compatible with other deep learning models. We use a pre-trained ResNet-34 from the Dlib library for facial landmark detection. The model is highly robust and is able to analyze non-frontal faces, as well as occluded landmarks relatively well.

From the detected facial landmarks over a series of frames, we can directly identify some facial activities such as yawning, napping, blinking, and talking. For yawning, we define this activity based on the distance among the lip landmarks. For a given frame sequence $[f_m,f_n]$, the facial landmark of upper lip and lower lip is $p_{ul}(f_i)$ and $p_{ll}(f_i)$ respectively, if the condition of $\mathcal D_W(p_{ul}(f_i), p_{ll}(f_i))>\mathcal D_{thrl}$ is met for all frames $f_i \in [f_m,f_n]$, the yawning activity is detected for this specific frame sequence. $\mathcal D_W(p_1(f_i),p_2(f_i))$ denotes the average distance of point $p_1(f_i)$ and $p_2(f_i)$ for a sliding window of $W$ frames.

Similarly, the facial landmarks of upper eyelid and lower eyelid is $p_{ue}(f_i)$ and $p_{le}(f_i)$. If the condition of $\mathcal D_W(p_{ue}(f_i), p_{le}(f_i))>\mathcal D_{thre}$ is met for frames $f_i \in [f_m,f_n]$, the eye-closing status is detected in this sequence. The threshold for the yawning $\mathcal D_{thrl}$ and eye closing $\mathcal D_{thre}$ is defined manually considering the difference of individual. Based on this, the facial activity is classified as napping when the time span is relatively long $f_n-f_m>f_{thr}$, otherwise the activity is classified as blinking. 

\begin{figure}[h]
	\begin{center}
		\includegraphics[width=0.45\textwidth]{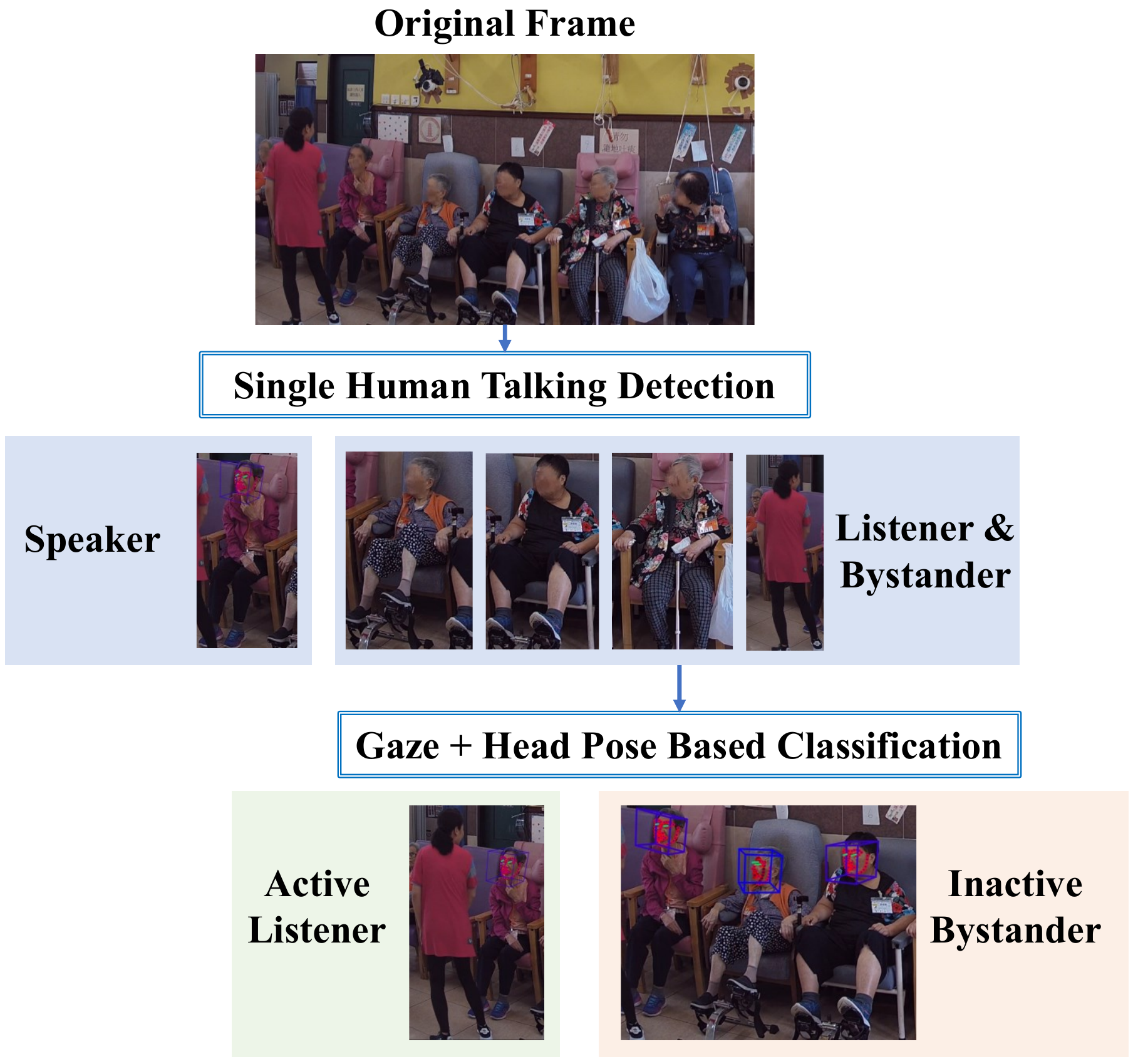}
	\end{center}
	\caption{Illustration of talking activity detection: a hierarchical two-stage classification paradigm. Faces are blurred for privacy protection.}
	\label{Figure:talking}
\end{figure}

Talking is different from other activities since it involves both the speaker and the listener. Fig.~\ref{Figure:talking} illustrates the framework of talking activity detection for which we adopt a hierarchical two-stage classification paradigm. Firstly, the talking behavior of an individual is detected based on the facial landmarks, then all human subjects are analyzed and in turn categorized as a speaker, or non-speaker. We then further classify the non-speaker into two fine-grained types: active listener or inactive bystander for which we use gaze and head pose detection to identify such talking interactions. The active listeners are those who the speaker is talking to and can be detected if their gaze directions pass through the head bounding box of the detected speaker. These inactive bystander include people listening but not interacting with speaker and people not listening at all.

\subsection{Activity Analysis}

Based on the suggestion from the caregivers, we automatically analyze a set of day-to-day activities of senior citizens. In particular, the care center routinely conducts guided group physical exercises, detailed analysis of which can gain good insights to the exercise intensity and fitness of the elderly. In light of this, we propose an action detection module consisting of two parts: exercise intensity detection which aims to monitor the activeness level during group exercises, and activity detection which focus more on the other non-trivial daily activities including interaction with other objects in the care center. 

\subsubsection{Exercise Intensity Detection}
\begin{figure}[h]
	\begin{center}
		\includegraphics[width=0.45\textwidth]{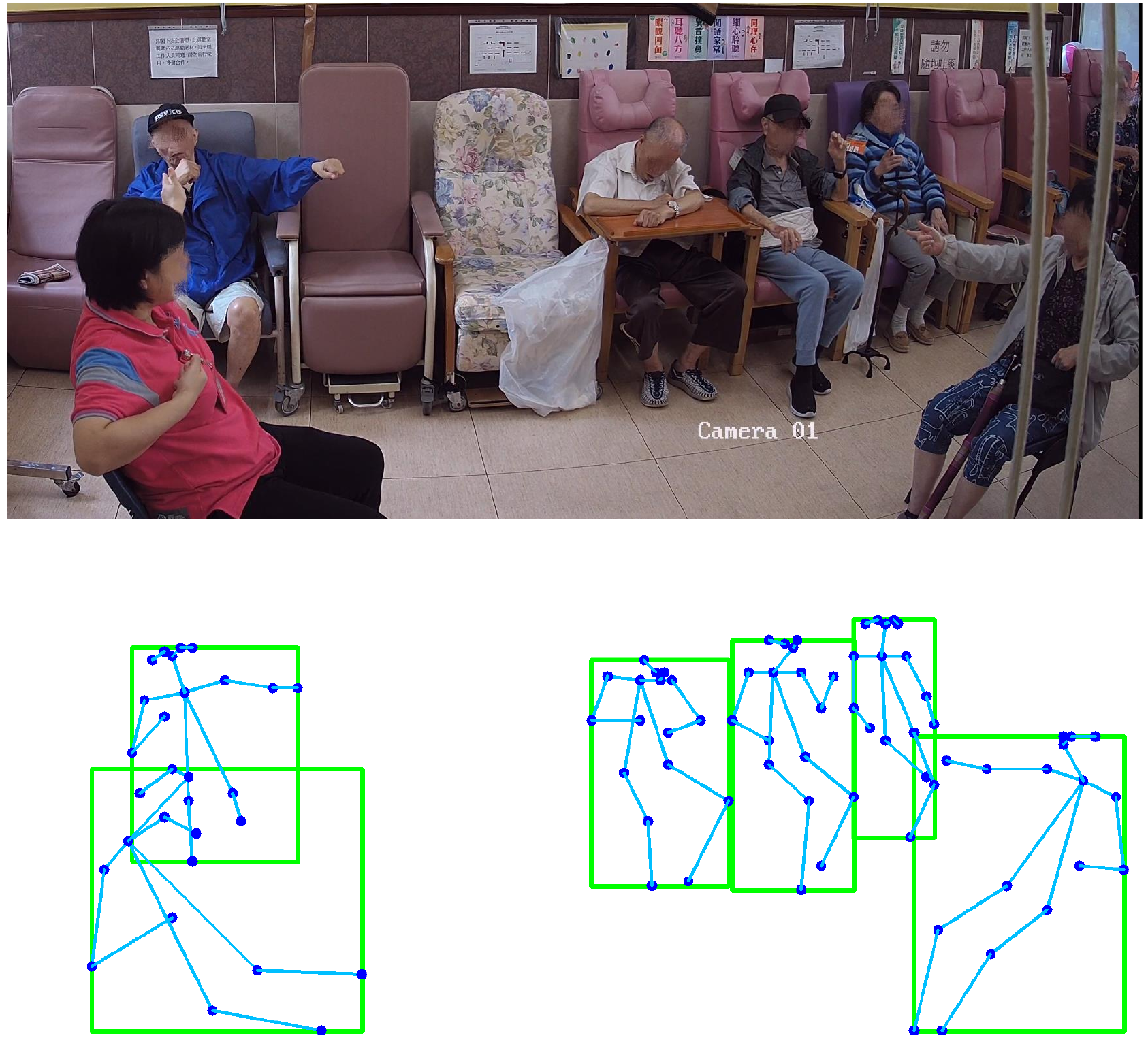}
	\end{center}
	\caption{Illustration of daily guided group exercises in the care center. We utilize pose features to derive the activeness level of the elderly. Faces are blurred for privacy protection.}
	\label{Figure:pose}
\end{figure}

In the care center, senior citizens must follow guided group exercises daily to maintain their physical fitness, as illustrated in Fig.~\ref{Figure:pose}. We attempt to automatically analyze each individual's exercise intensity in such group exercises which should reflect their physical health conditions. We define a metric \emph{exercise intensity score} which measures the activeness level in the guided group exercise. We first extract the key points of each person's 2D body pose, obtained using OpenPose~\cite{8765346}. To compute \emph{exercise intensity score} from these extracted points and their dynamics entailing various aspects of their body movements, we derive three meta features: \emph{angle}, \emph{range}, and \emph{speed} of body movements.

\begin{figure}[h]
	\begin{center}
		\includegraphics[width=0.45\textwidth]{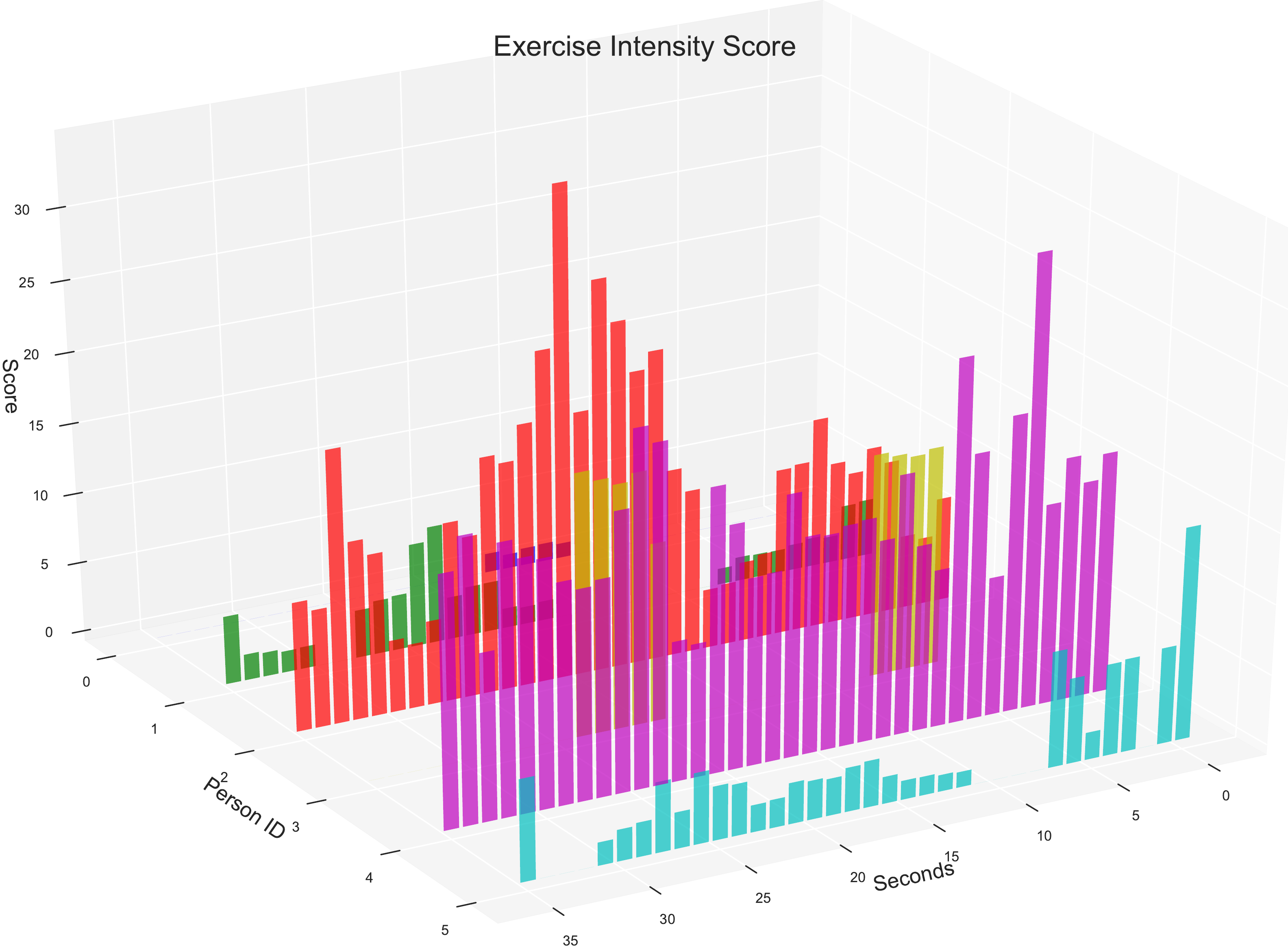}
	\end{center}
	\caption{Exercise intensity score for different senior citizens within 35s. We can see from the bar that elderly \#2 and \#4 actively were doing exercise with higher intensity.}
	\label{Figure:excercise-score}
\end{figure}

Based on the elbow keypoints obtained from pose estimation of a frame over a sequence of frames, we compute \emph{angle} via $arccos$($l_{se} \cdot l_{ew} / |l_{se}| |l_{ew}|$), where $l_{se}$ and $l_{ew}$ represent the keypoint vector from the shoulder to the elbow, and from the elbow to the wrist respectively. Then we normalize \emph{angle} to $[0,1]$ to get the angle score as $f_{\emph {angle}}$. The \emph{angle} encode the pose feature and can guide the inference of body movement during exercise. Similarly, we compute the movement range of wrist keypoint and normalize it to $[0,1]$ to derive $f_{\emph {range}}$, and compute normalized $f_{\emph {speed}}$ considering the time taken to reach the maximum body movement. We combine these three meta features with the weighting coefficients $\lambda_{a}, \lambda_{r}, \lambda_{s}$ to balance between the different components. Fig.~\ref{Figure:excercise-score} illustrates the 3D box plot of how \emph{exercise intensity score} $f_{\emph {eis}}$ varies among different senior citizens.  

\begin{equation}
    f_{\emph {eis}} = \lambda_{a} \cdot f_{\emph {angle}} + \lambda_{r} \cdot f_{\emph {range}} + \lambda_{s} \cdot f_{\emph {speed}}
\end{equation}

%\section{System}
\subsubsection{Temporal Activity Detection}
In addition to detecting the activity level during the guided group exercises, we detect various activities, including sitting, watching TV, eating, drinking, and doing individual exercise. We can define our temporal activity classification as a $K$-way classification ($K=5$ for our system) with a training set $D_{train}={\{(x_i,y_i)\}_{i=1}}$, where $x_{i} \in \mathbb{R}^{T \times H \times W \times C}$ is the video input and $y_{i} \in [0,1]^{K}$ is a vector representing whether an activity from the whole $K$ defined activities is performed or not in this video. Our goal is to train a model learning a mapping $f:x_{i}\xrightarrow{} y_{i}$ from video input to the activity vector label. This can be done by minimizing the loss function:
\begin{equation}
    f^{*}=\mathop{\arg\max}_{f \in \mathcal{F}} \frac{1}{|D_{train}|} \sum_{(x_i,y_i) \in D_{train}} \mathbb{E}[l(f(x_{i}), y_{i})],
\end{equation}

where $\mathcal{F}$ is a class of functions and $f(x_{i})$ is the softmax score representing the probability of each category. $l$ is the loss function and we use softmax cross entropy loss:
\begin{equation}
l(f(x_{i}), y_{i}) = -\sum_{j=1}^{K} \mathbf{1}(y_{i}=j)\text{log} f_{j}(x_i),
\end{equation}
where $\textbf{1}$ is an indicator function that equals 1 when condition is true and 0 when false. Once we have a pre-trained model $f$, we also want to fully utilize the temporal information for better accuracy. Given a continuous video $\mathcal{V}$, we segment it into $L$ overlapping clips of $T$ frames $X=\{x_i\}_{i=1}^{L}$, where $x_{i} \in \mathbb{R}^{T \times H \times W \times C}$. Each video clip is input into our model $f$ and we use a sliding window $W$ to smooth the predictions:
\begin{equation}
    \bar{f}(x_i)=\frac{1}{W} \sum_{j=i-W/2}^{j=i+W/2} f(x_j).
\end{equation}

Additionally, for different activity categories, the best sliding window length $W$ is different, which is different from using a fixed window length. We apply grid search on different activities and find the best sliding window length. In the experiment part, we will show the precision improvement boosted by the sliding is significant.

\subsection{Interaction Analysis}

Some activities involve the interaction between a person and a certain object. These human-object interaction includes sitting on chair, watching TV, eating at table, and drinking with bottle or cup. While our temporal activity detection module is capable of predicting which activity is performed in a given frame sequence (e.g. predicting "drinking"), we also want to locate the position of human and object and assign the detected interaction to a pair of detected human instance and object instance (e.g. detecting "Senior \#1 is drinking with a bottle"). 

The task of detecting these activities can be formulated as detecting a triplet $\langle human, verb, object \rangle$. Similar to previous human-object interaction detection model such as InteractNet~\cite{gkioxari2018detecting}, we build a network consists of three network branches: human, object, and interaction. The object detection branch built on Fast R-CNN~\cite{ren2015faster} is capable of detecting candidate human/object boxes $b_h$,$b_o$ and predict the class score $s_h$,$s_o$. Then the cropped human instance is input to the human branch, we extract feature with RoIAlign~\cite{he2017mask} and predict a score $s_h^a$ for each action $a$. The interaction branch receive the input of both cropped human and object boxes $b_h$,$b_o$. Then we predict probabilities $s_{h,o}^a$ for each action $a$. The output layer for human and interaction branch consists of four binary sigmoid classifiers for \textit{multi-label} classification. The final prediction for each interaction $a$ is
\begin{equation}
    \mathcal{S}^a = s_h \cdot s_o \cdot s_h^a \cdot s_{h,o}^a .
\end{equation}
The training objective is to minimize the binary cross-entropy between the ground-truth action labels and scores $\mathcal{S}^a$ predicted by our model.

Besides interaction of human and objects, human-human interaction is another important analysis target of a healthcare system. As we have built the facial analysis and temporal activity detection module, we can get the clues of social interaction from the detected human-human interaction - talking activity. Our intuition is that the more talkative an individual is, the better mental health condition he or she is in. This association has been proved in previous research, especially for dementia patients~\cite{rousseaux2010analysis, adams2005communication}. 

\begin{figure}[h]
	\begin{center}
		\includegraphics[width=0.45\textwidth]{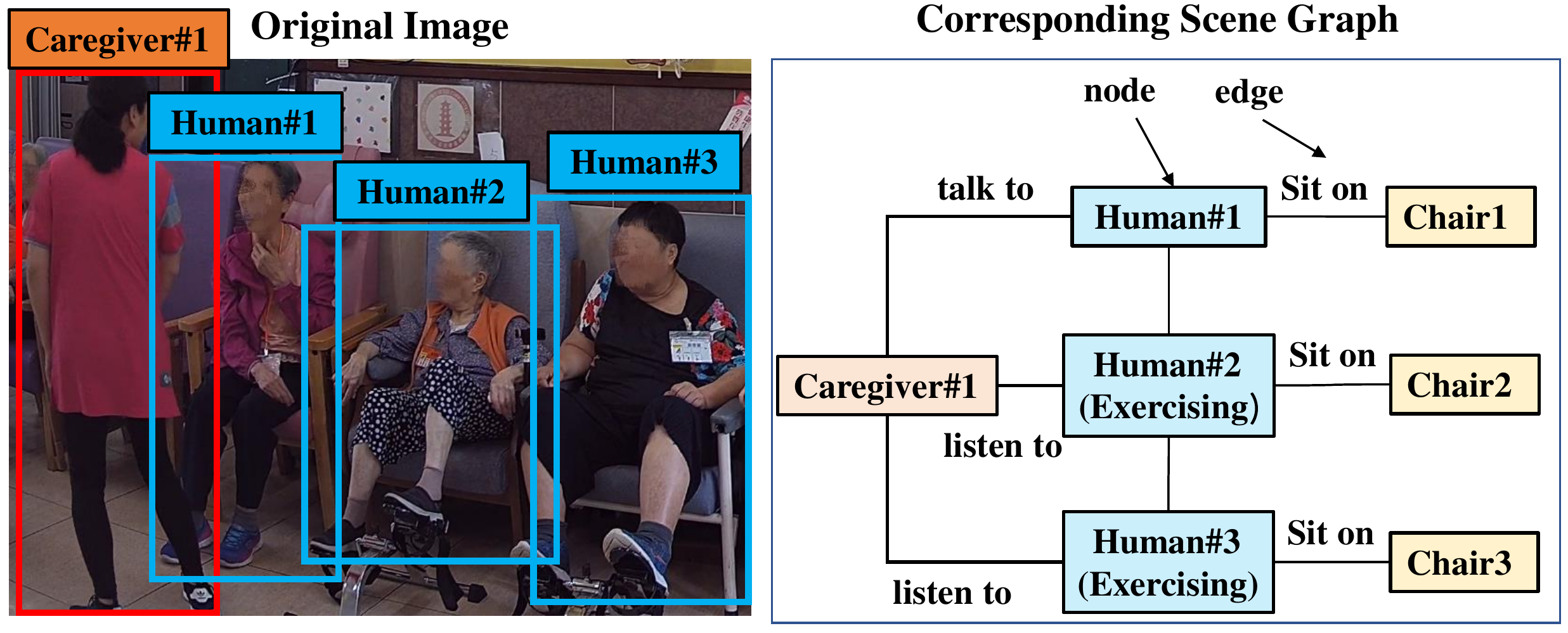}
	\end{center}
	\caption{An example of scene graph: nodes representing human and object instance and edges representing the relationship between two nodes.}
	\label{Figure:sg}
\end{figure}

We adopted \emph{scene graph} to visualize and describe the activities in a given scene in a straightforward manner. \emph{Scene graph} was first proposed by Johnson~\textit{et~al.}. as a data structure. Formally, scene graph $G=(O,E)$ is a directed graph, where $O= o_1, ... ,o_n$ is the detected instance in the images or the frame of a video. Each detected instance $o_i=(c_i, A_i)$ has category $c_i$ and attribute $A_i$ (e.g. a human is doing exercise). $E$ is a set of directed edges that represent the relationships between objects. Fig.~\ref{Figure:sg} gives an example of a scene graph describing a given frame of our video recording data. We can see that the scene graph contains different nodes representing the senior citizens, caregivers, and objects such as chairs. There are also edges representing the relationship between nodes. 

To understand the mental wellness of senior citizens, we define a feature for the elderly node in the graph. The feature can be computed via aggregating the information of neighboring nodes and their edges. The most important is the talking activity between human nodes. We give this activity high weight as we believe a talkative person has a better mental health condition. There are also other activities such as watching TV. These activity has some contribution to the mental wellness but not significant. 

Based on the scene graph, we define the \textbf{close relationship} between two persons. If the average talking time between the pair is more than a given threshold $t_{th}$, we say that the pair of persons has a close relationship. Note that the pair of persons can be two senior citizens, or one caregiver and one senior citizen. The mental wellness system can provide guidance for possible psychotherapy accordingly.

\subsection{Analysis of Long-term Pattern and Trend}

\begin{figure*}[h]
	\begin{center}
		\includegraphics[width=\textwidth]{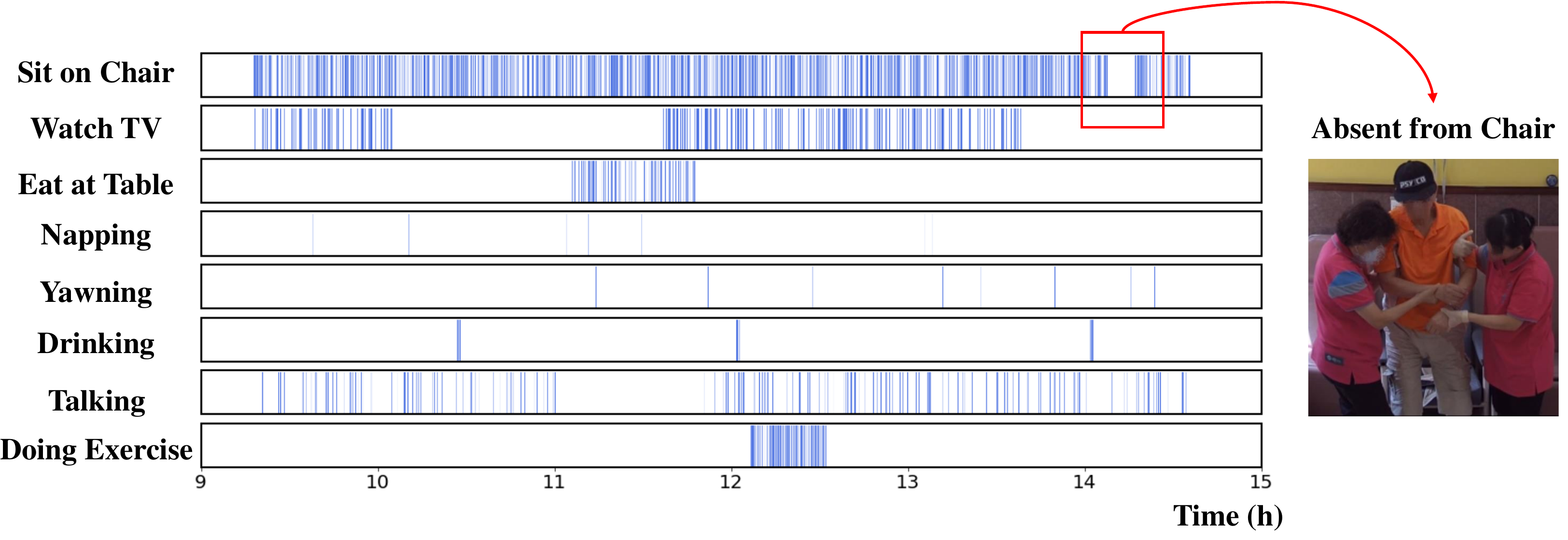}
	\end{center}
	\caption{Activity temporal heatmap of eight activities in a given day. The dark blue at certain time indicate higher probability of an action was being performed. Red box denotes the missing of ``sit on chair'' action and our system will alert caregivers accordingly. }
	\label{Figure:heatmap}
\end{figure*}

Fig.~\ref{Figure:heatmap} gives an example of \emph{activity temporal heatmap} visualizing the activity of a specific senior citizen during a whole day. Compared to the probability prediction of various activities, the heatmap provides a more straightforward illustration of what and when an activity is performed. In the example figure, we can see that the elderly citizen was sitting on the chair all day, except for some time slot. The system can alert the caregivers when the elderly was not on the chair. Additionally, we can see which time during the day a senior citizen is more likely to be sleepy and change the group exercise time accordingly.

\begin{figure}[h]
	\begin{center}
		\includegraphics[width=0.48\textwidth]{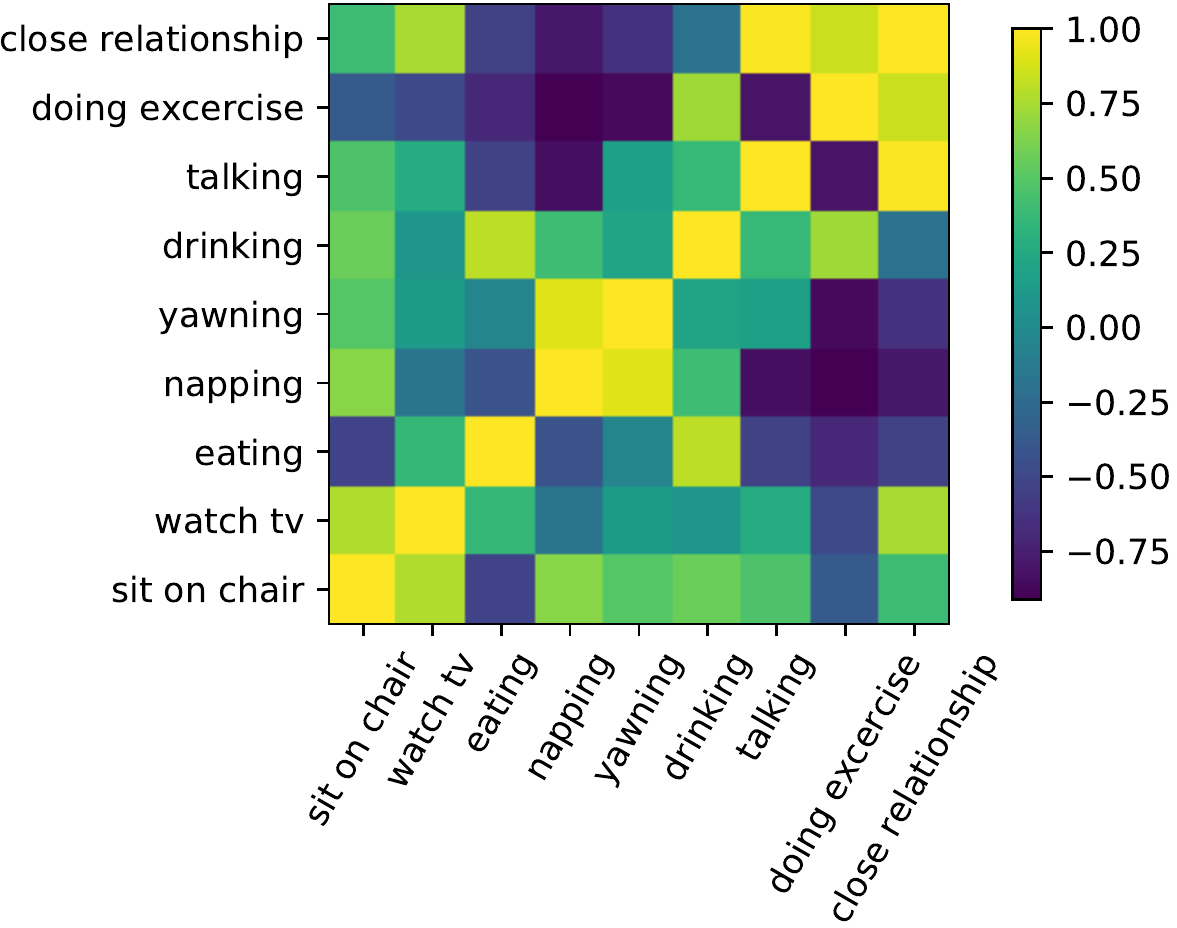}
	\end{center}
	\caption{Correlation Matrix of different activities. Positive correlation implies the two actions are more likely to be performed together.}
	\label{Figure:corr}
\end{figure}

Some actions are more likely to be performed together while others not. To mine the association rule in our detection result, we use correlation to measure the co-occurrence. The correlation $\eta$ is defined as:
\begin{equation}
    \eta = \frac{cov(X,Y)}{\sigma_X \sigma_Y} = \frac{E((X-\mu_X)(Y-\mu_Y))}{\sigma_X \sigma_Y},
\end{equation}
where $X$ and $Y$ are probability distribution output of two activities, $E(X)$ is the expected value of distribution $X$ and $\sigma_X$ is the variance of distribution $X$. We compute the correlation between each activity pairs and show them in the co-occurrence matrix in Fig,~\ref{Figure:corr}.

Both intuitive and some counter-intuitive results can be observed. For example, some activities, such as yawning and napping, eating, and drinking are more likely to be performed together. While watching TV and napping are highly exclusive. This is close to our common sense. There are also interesting findings that may provide insights for the caregivers. As an instance, if someone watches TV more, he or she is more likely to have more close relationship with others. These correlations can help improve the accuracy of prediction via the following prediction correction inspired by~\cite{benezeth2009abnormal}:
\begin{equation}
    p_{a_i} \xleftarrow{} p_{a_i} + \sum_{a_j \in A, a_j \neq a_i } \eta_{ij}p_{a_j}
\end{equation}
For activity $a_i$ in the action set $A$, we use the correlation $\eta_{ij}$ to correct the prediction $p_{a_i}$ based on the activity prediction $p_{a_j}$. This feedback paradigm based on correlation analysis effectively boosts the performance of our model and we will show the effectiveness in the experiment part.

\begin{figure}[h]
	\begin{center}
		\includegraphics[width=0.5\textwidth]{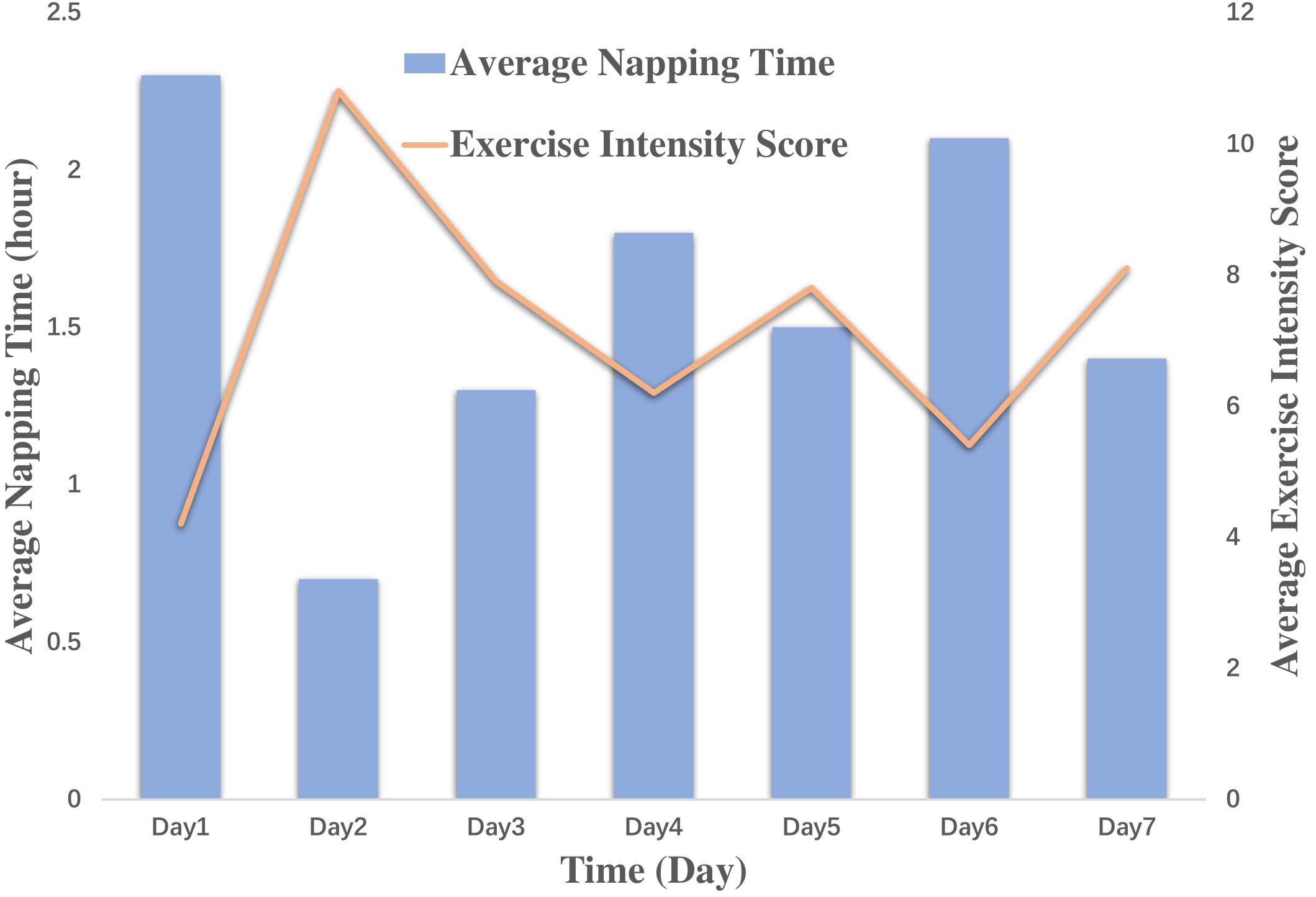}
	\end{center}
	\caption{Average napping time and exercise intensity score in consequent seven days. We can see the tendency of the napping time and obvious negative correlation between the napping time and exercise intensity score.}
	\label{Figure:napp}
\end{figure}

%%%%%%%%%%%%%%%%%%%
% We analyze the long-term analysis and found xxx yyy zzz. 
%%%%%%%%%%%%%%%%%%%
Besides the correlation analysis, we also inspect the activities and features in the long term. Fig.~\ref{Figure:napp} gives an example of the activity timeline of napping and the average exercise intensity score during the group exercise in 7 days. It allows caregivers to observe the changes of specific features and can alert the caregivers when there is an anomalous increase or decrease of some metric, e.g. too long napping or lower exercise intensity than normal.

\section{Evaluation} 
%%% this part should be the meat of the paper but the content here is very small compared to the rest. %%%%%%%%%%%%%
In this section we show how to collect and process video recording dataset. To extensively evaluate the system, we need some corresponding label to check if our prediction was correct. One of the most representative and important module in our system is the activity detection module. We choose the activity detection task for evaluation. 

\subsection{Data Collection}
Our dataset consists of numerous videos, each of which contains elderly individuals doing their daily activities at an elderly care center. This dataset was obtained from the Haven of Hope Elderly Homes, which is a care center providing holistic care for frail elders and chronically ill patients. To gather the dataset, we have obtained permissions from the Haven of Hope Elderly Homes to record their citizens’ daily activities within the working hour periods. These activities consist of both individual activities such as watching television, individual physical exercise, eating – and group activities such as interaction with other citizens, interaction with the caregivers, aided group physical exercise.  These videos are recorded at 25FPS, 4K resolution which allowed for more accurate inference and analysis to be performed. Overall, we have \textbf{168} hours of video recording of four cameras from different angles. There are a total \textbf{60M} frames with 4K spatial resolution. The recording time span a whole month which can provide research possibility for long-term feature analysis. We only use our data for research purposes and blur all the faces when showing results.

\begin{table*}[h]
\centering
\resizebox{\textwidth}{!}{
\begin{tabular}{l| c c c c c c c|c}
\hline
Average Precision (AP)  & sit on chair & watch tv & eating & drinking & napping & yawning & talking & mAP \\
\hline
\hline
    VPN~\cite{das2020vpn}                                & 0.844 & \textbf{1.000} & 0.769 &0.896 &- & - & - & -\\
    I3D~\cite{carreira2017quo}                                & 0.792 & 0.920 & 0.722 &0.858 &- & - & - & -\\
    Active Speakers Context~\cite{alcazar2020active}            & - & - & - & - & - & - & 0.689 & -\\
\hline
    baseline                           & 0.604 & 0.893 & 0.735 &0.829 &0.374 & 0.712 &0.546 & 0.670\\
    baseline + fixed sliding           & 0.773 & 0.893 & 0.687 &0.844 &0.412 & 0.890 &0.727 & 0.747\\
    baseline + sliding                 & 0.858 & 0.893 & 0.714 &0.844 &0.581 & 1.000 &0.727 & 0.802\\
    baseline + co-occurrence           & 0.693 & 0.946 & \textbf{0.782} &0.851 &0.443 & 0.756 &0.746 & 0.745\\
    baseline + co-occurrence + sliding & \textbf{0.862} & \textbf{1.000} & 0.769 & \textbf{0.902} &\textbf{0.650} & \textbf{1.000} & \textbf{0.746} & \textbf{0.847}\\
\hline
\end{tabular}}
\caption{Activity classification result of Average Precision (AP) comparison on our test set. VPN~\cite{das2020vpn} and I3D~\cite{carreira2017quo} are trained on Kinetics~\cite{kay2017kinetics} and Toyota Smarthome~\cite{dai2020toyota} action classification dataset. Active Speakers in Context~\cite{alcazar2020active} is trained on AVA-Active Speaker dataset~\cite{roth2020ava}. }
\label{tab:ap}
\end{table*}

\subsection{Results}
We have manage to provide the healthcare assistance for 21 senior citizens in the care center. Monitoring the health condition of all the 21 senior citizens during the day time is not feasible for caregivers. These long-term feature and trend also cannot be simply observed by caregivers. With the help of our system, these challenges are tackled.

We select the activity classification task for evaluation task and mean Average Precision (mAP) as the evaluation metric. To evaluate the performance of our the activity detection module, we select some representative clips from the whole dataset and label the instance-level activity. There are totally more than 4000 instances in the test set, covering seven activities excluded exercising (because exercising time is fixed every day in our data and we manually set a window to filter out all the false positive detection). We have made certain efforts, including dynamic sliding window length, correcting the prediction with activity co-occurrence. 

Tab.~\ref{tab:ap} shows the Average Precision (AP) of the instance-level activity classification. We define a detection as correct when the $K$ ground truth activities are the same as top-$K$ predicted activity categories. The ``fixed sliding'' indicates the sliding window with fixed length and ``sliding'' in the table represents the average sliding with the dynamic length for different activities. The ``co-occurrence'' implies the prediction correction based on the correlation. We can see that both sliding window and co-occurrence correction boost the performance of our model on the test set. Compared to the baseline model, the sliding windows and correlation correction improves the mean average precision (mAP) by 19.7\% and 11.2\% respectively. Combination of all tricks boost mAP by 26.4\%.

Compared to state-of-the-art activity classification model VPN~\cite{das2020vpn} and I3D~\cite{carreira2017quo}, our model achieves better accuracy on four activity categories. As baseline model fails to outperform VPN and I3D, we can see the significant contribution of temporal information and activity correlation. For talking detection, we adopt a facial landmark based solution and it surprisingly outperforms state-of-the-art multi-modal speaker detection model Active Speakers in Context~\cite{alcazar2020active}. This is mainly because our video recording has poor audio quality and multi-modal methods fail to utilize the acoustic feature.

\subsection{Technical Challenges} 

\begin{figure}[h]
     \centering
     \begin{subfigure}[b]{0.15\textwidth}
         \centering
         \includegraphics[width=\textwidth]{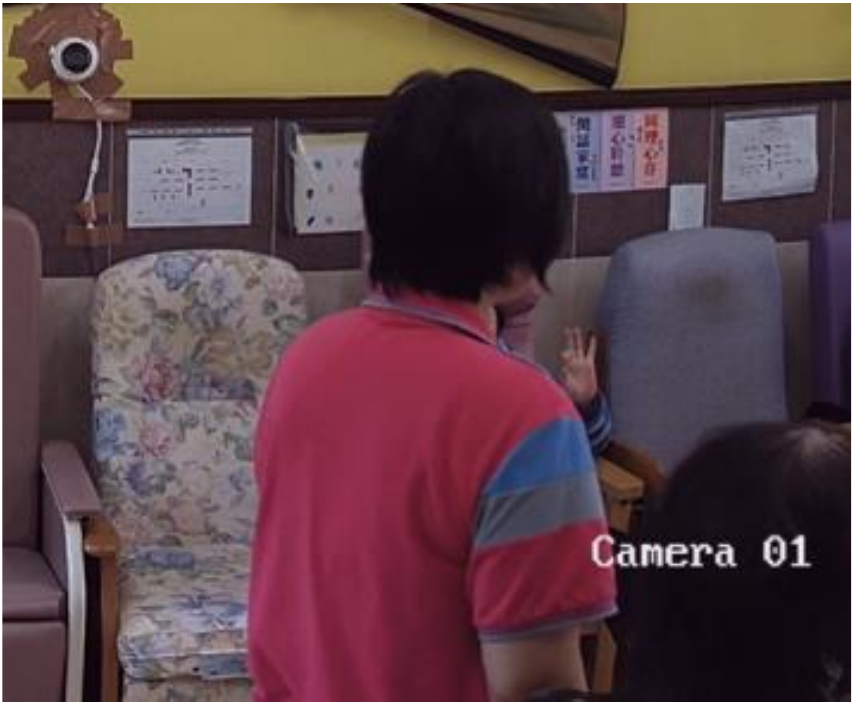}
         \caption{Original Image}
         \label{fig:y equals x}
     \end{subfigure}
     \begin{subfigure}[b]{0.15\textwidth}
         \centering
         \includegraphics[width=\textwidth]{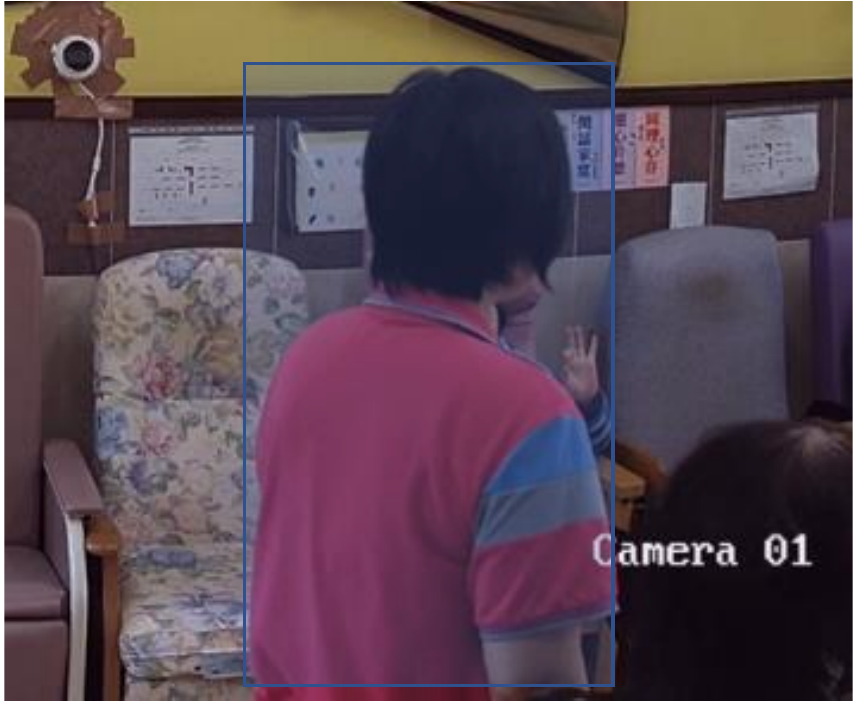}
         \caption{Detection Result}
         \label{fig:three sin x}
     \end{subfigure}
     \begin{subfigure}[b]{0.15\textwidth}
         \centering
         \includegraphics[width=\textwidth]{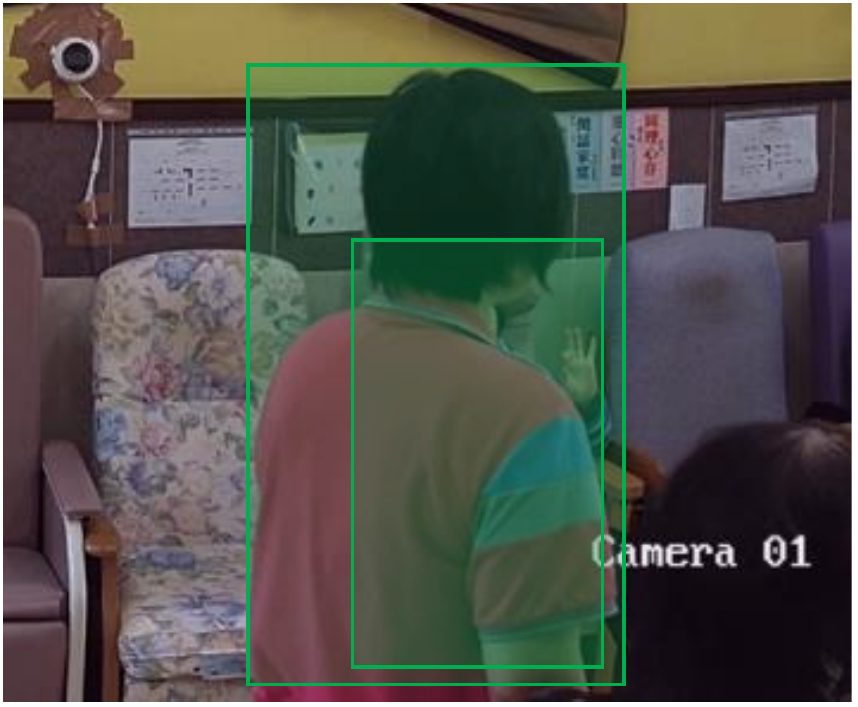}
         \caption{Ground Truth}
         \label{fig:five over x}
     \end{subfigure}
        \caption{An example of the detection error.}
        \label{fig:occlusion error}
\end{figure}

In this section, we will talk about the technical challenges in our system and possible technical enhancement for an accurate and fully automated solution. The first core challenge is the accuracy of our system. Since we are not using an end-to-end pipeline, the detection and recognition error in each stage of the system will accumulate to the end. An example of early stage detection error is given in Fig.~\ref{fig:occlusion error}. Besides the occlusion, the motion blur, video defocus, and pose changes all pose challenges for more accurate detection. The difference between public dataset and our video dataset also drop the accuracy for activity analysis module.

The development of a deep neural network makes it possible for more robust and accurate detection in video. We are also considering incorporate a better temporal model to replace the simple sliding average. However, too complicated network architecture will bring to another problem: efficiency and computational cost. We need to strike a balance between the accuracy and effectiveness.

For the interaction part, we do not incorporate the detection of some subtle interactions, such as ``help someone''. The problem is that the definition of these interaction are vague and it is hard to transfer from one data set to another. We also choose not to incorporate the detection of some subtle facial expression in our system. These facial expressions include disgust, anger, sad, and so on. Their detection will be very informative for understanding the mood and mental state of senior citizens~\cite{fei2020deep}. While there are some pre-trained model applicable for the detection, the performance on our senior healthcare application is limited because the facial expression of these senior citizens is not active enough compared to younger people~\cite{asgarian2019limitations,kunz2017problems}. One possible solution is to finetune the facial landmark model more on the data specifically collected for elderly healthcare application.

\section{Conclusion}
% There is an 8-page limit.

We build an automated vision-based wellness analysis system and demonstrate its effectiveness on the healthcare of elderly care center citizens. Additionally, we extensively perform long-term feature analysis which can be a reference for the caregivers.
We are still working on the project to incorporate more components into our model. 
This includes, but not limited to, more long-term analysis, personalized healthcare profiles, and more activities including abnormal ones.
Our final goal is to build an \textbf{privacy-preserving, robust, and high-efficiency} healthcare system that provides comprehensive to both the caregivers and elderly without high computation cost.

\newpage
\bibliography{aaai22.bib}

\clearpage
\appendix

\section{Appendix}

\subsection{Data Labeling}

\begin{figure}[h]
	\begin{center}
		\includegraphics[width=0.4\textwidth]{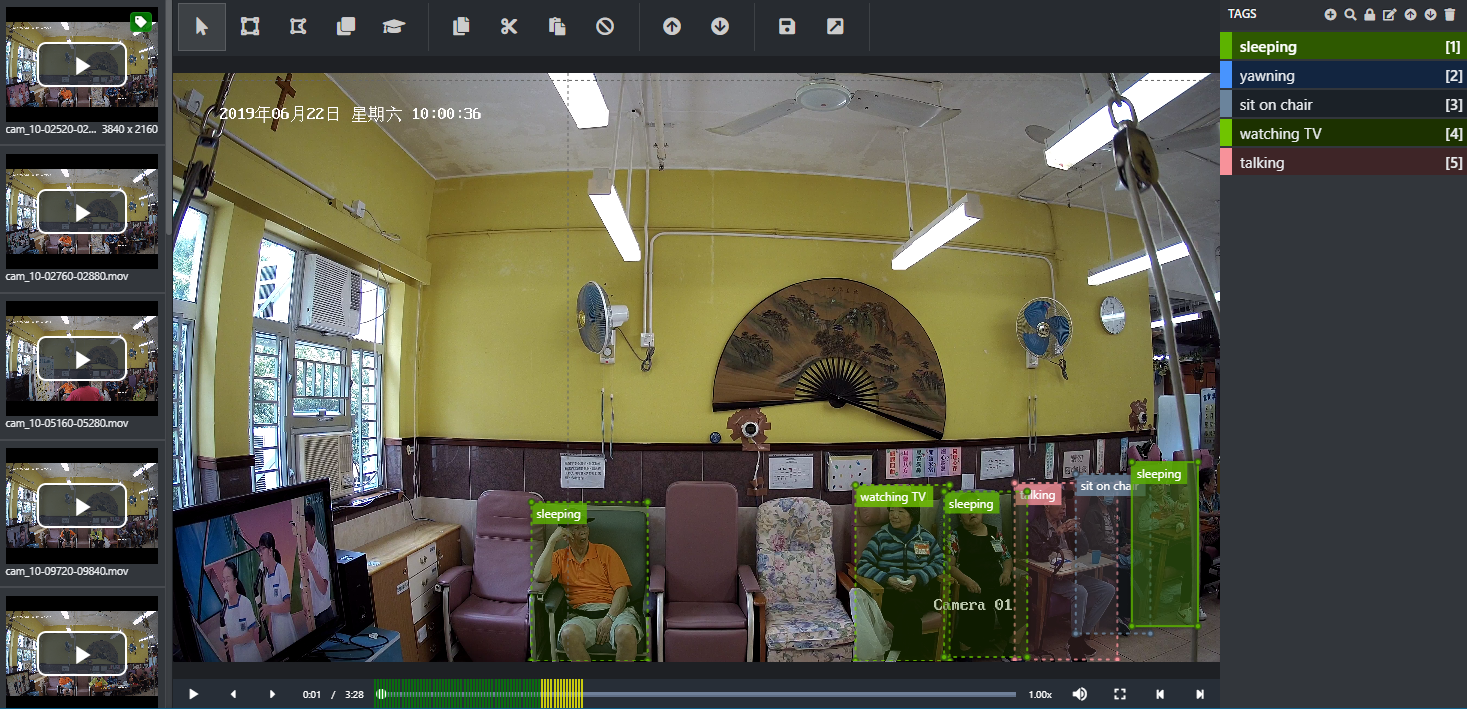}
	\end{center}
	\caption{Data label tool user interface.}
	\label{Figure:label}
\end{figure}

\begin{figure}[h]
	\begin{center}
		\includegraphics[width=0.4\textwidth]{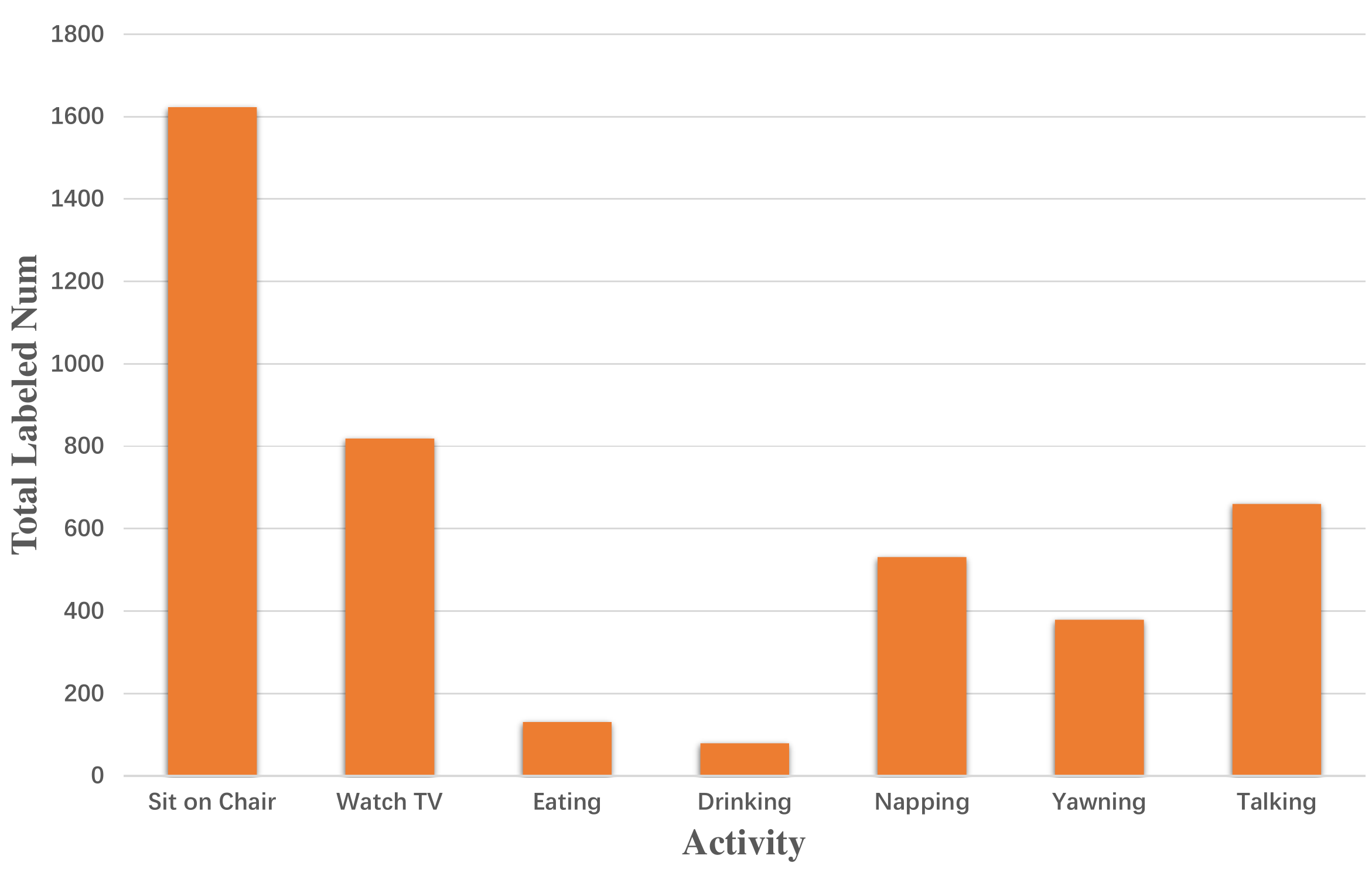}
	\end{center}
	\caption{Data label distribution of our test set.}
	\label{Figure:datadis}
\end{figure}

After collecting high-quality video from the elderly care center, we label the activity in instance-level using open-source annotation and labeling tool  VoTT (Visual Object Tagging Tool)~\footnote{https://github.com/microsoft/VoTT}. The user interface of VoTT is shown in Fig.~\ref{Figure:label}. For each given frame, we exhaustively label all activity for each human in the video. Fig.~\ref{Figure:datadis} gives an analysis of label distribution. We can see from the statistic that our selected video clip covered all activity and reflect the real distribution of them.

\subsection{Insights from exercise intensity score result}
\begin{figure*}[h]
	\begin{center}
		\includegraphics[width=0.9\textwidth]{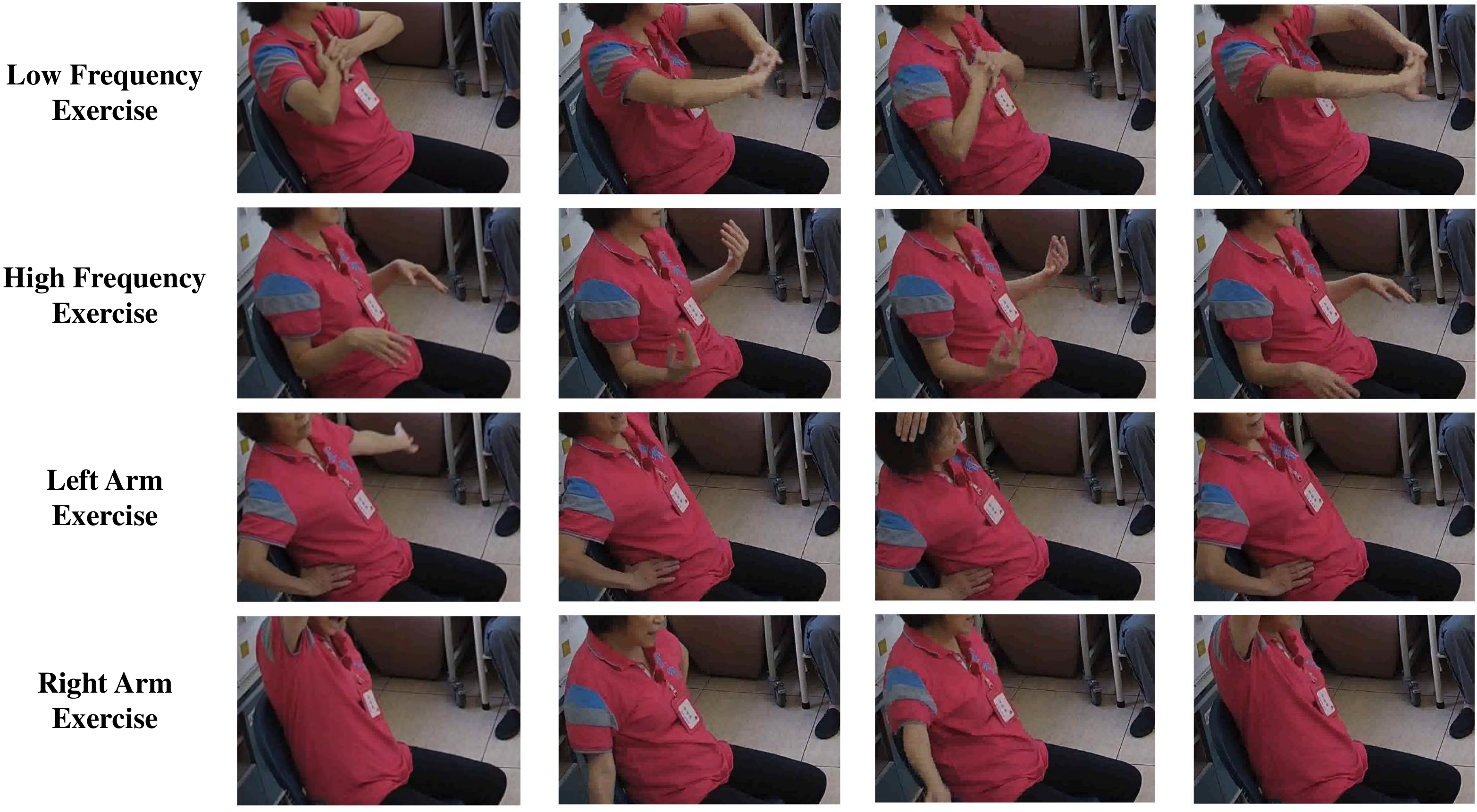}
	\end{center}
	\caption{An example of four exercise categories.}
	\label{Figure:exc}
\end{figure*}

Considering the characteristic of group exercise, we categorize them into four subsets: low frequency exercise, high frequency exercise, left arm exercise, and right arm exercise. An example of these exercise types is shown in Fig.~\ref{Figure:exc}. Guided group exercises are designed for senior citizens to keep them physically active and we hope senior citizens get comprehensive exercise for all body parts with proper frequency. Too much high-frequency exercise increase the risk of potential injuries while only low-frequency exercise has little effect on fitness keeping. Also, we wish seniors have balanced exercise on both left and right.

\begin{figure*}[h]
	\begin{center}
		\includegraphics[width=0.95\textwidth]{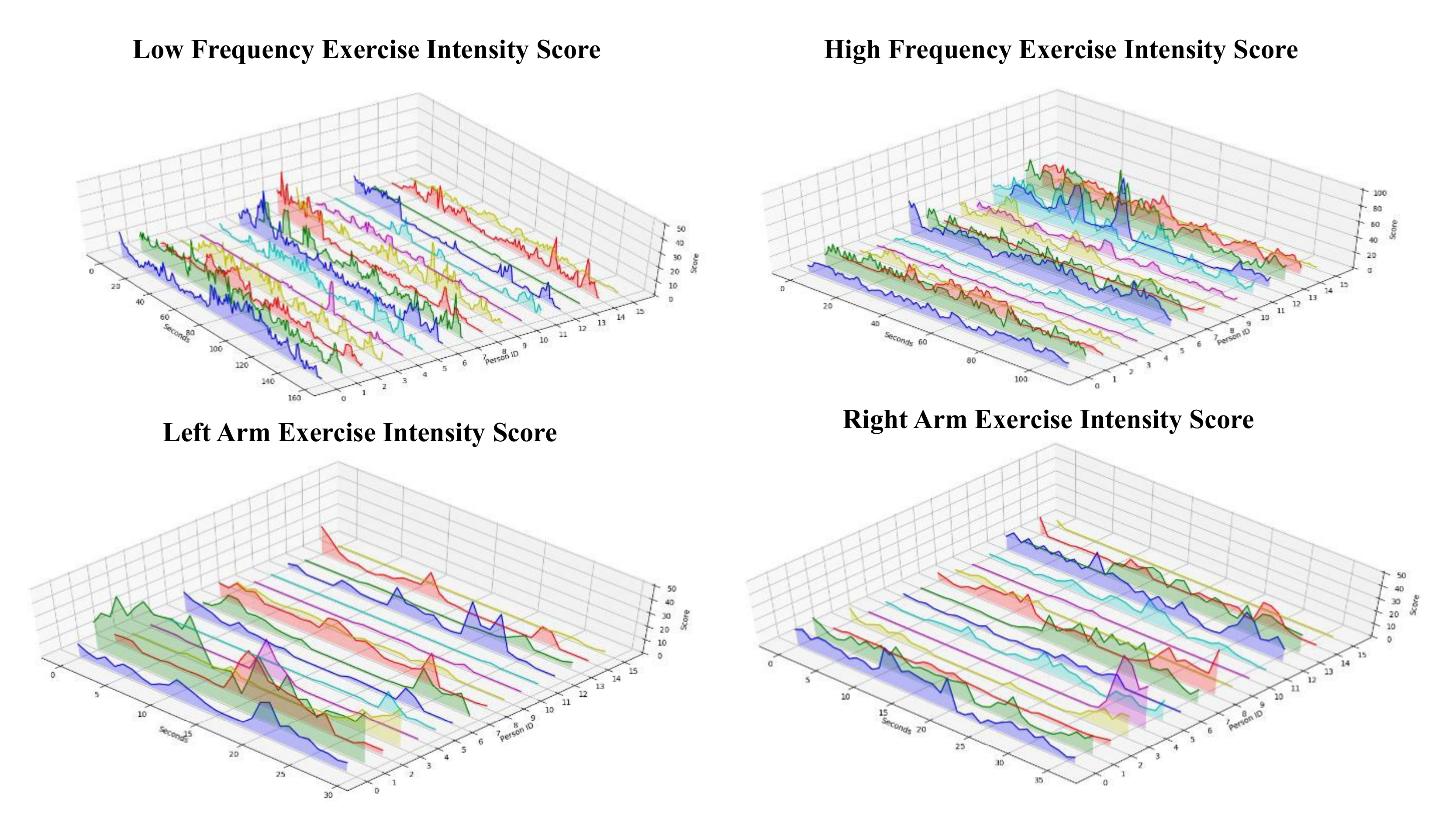}
	\end{center}
	\caption{Exercise intensity score of four exercise categories.}
	\label{Figure:4cls}
\end{figure*}

The exercise intensity score of the individuals are recorded during a whole week. Each day consists of 30-minute group exercise and 16 individuals including the reference individual or the caregiver (individual 0). Fig.~\ref{Figure:4cls} shows the exercise intensity score graph over time within the first 10 minutes of the group exercise. They are plotted according to the four exercise categories. In the figures, we can see the various intensity scores breakdown for each individual at a certain time.

In the low frequency exercise, the reference actor had a medium intensity score. This means that the majority of other individuals were also able to follow the exercise, shown in their medium intensity scores as well. There are few individuals who did not follow the group exercise, however. Individual 4, 10, 13 and 15 has significantly lower scores than the reference individual 0. This gives the insight that further assistance is needed for the individuals to do the same level of physical exercise. In doing so, caretakers may take action to give a more personalized physical exercise, or taking care of the relatively passive individuals’ needs. As a result, caregivers can allocate their caregiving attention batter.

Similarly for the high frequency exercise, we can draw observations that individual 12 has a higher intensity score relative to individual 0 in the first part of the exercise, but has very little to none score on the remaining duration of the exercise. From here, we can infer that individual 12 may needs further assistance in terms of extra rest, food, or other physical assistance. Caretakers will then be also able to provide a personalized physical exercise for the said individual to achieve similar levels of physical activity missed.

\begin{figure*}[h]
	\begin{center}
		\includegraphics[width=0.9\textwidth]{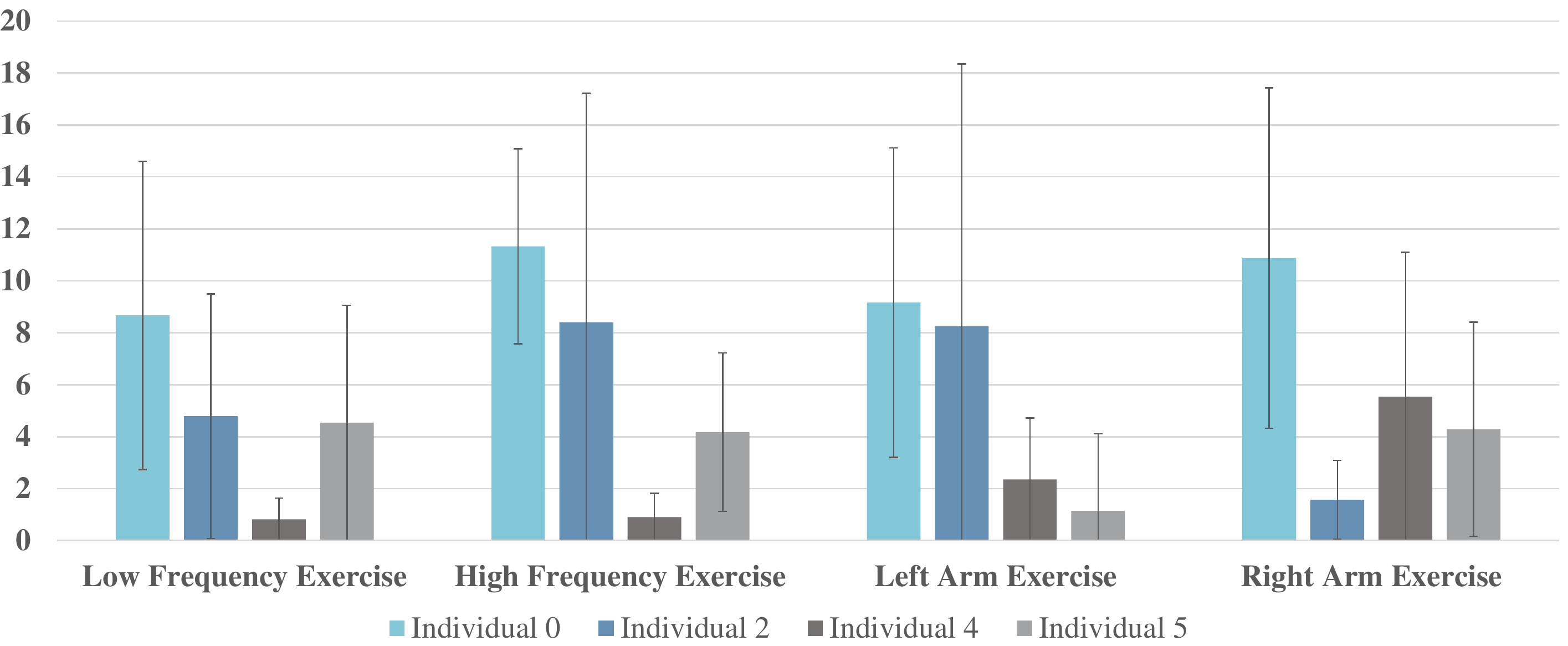}
	\end{center}
	\caption{Exercise intensity score analysis of four selected individual including one caregiver (individual 0) and three senior citizens (individual 2,4,5).}
	\label{Figure:eis}
\end{figure*}

The Fig.~\ref{Figure:eis} shows the mean and standard deviation of the exercise intensity scores four individuals within the span of four exercises during a day. According to the graph, individual 0 has the highest intensity score among the four individuals. This is denoted by the scores with mean values over 8. Meanwhile, individual 4 has the lowest intensity scores among the four individuals. This is denoted by the mean values of less than 5 and near to 0. On the other hand, individual 2 and 5 have a relatively medium level intensity score, as they have mean score values from 1 to above 8.

From this observation, it can be inferred that in terms of the group exercise, individual 0 is the most active. Individual 2 did the exercise as active as well, but did not follow the right arm exercise. Similarly, individual 5 also did not follow the left arm exercise. This observation is particularly very useful because exact ratios of the intensity can be drawn. For example, Individual 4 did the left arm exercise 25.76\% as intense as individual 0 in terms of the score mean. As such, caretakers may draw insights from this observation to give a more personalized and lighter physical exercise to the individual to achieve the same level of missing intensity score.

In conclusion, the observations taken from the physical intensity scores can provide new insights to the caretakers in taking the most suitable decisions to accommodate the patients. The physical intensity scores provide an overview of the individuals who are following the group exercise along with their intensity as well as the individuals who are not following the group exercise at all. 

\subsection{Insights from facial analysis result}

\begin{figure*}[h]
	\begin{center}
		\includegraphics[width=0.9\textwidth]{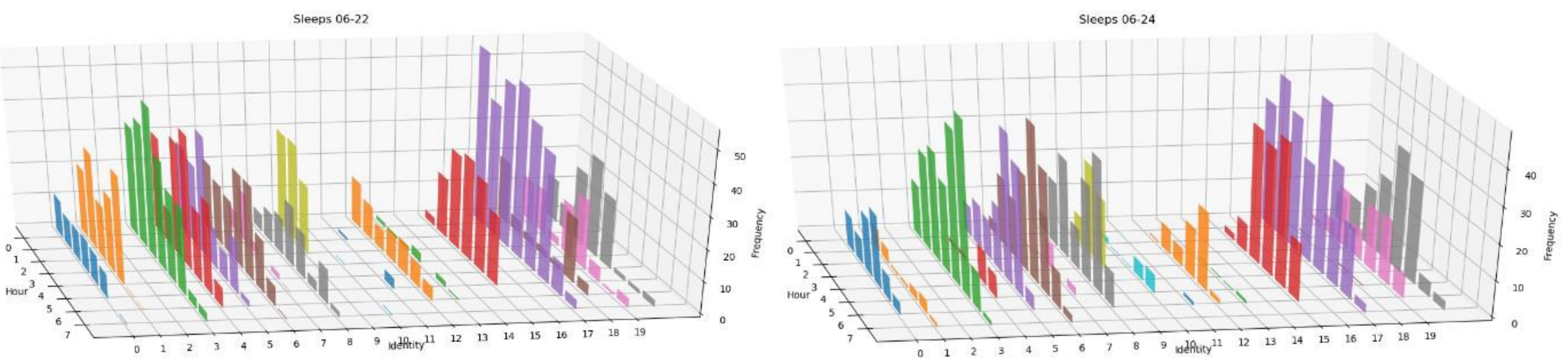}
	\end{center}
	\caption{Comparison of Napping time.}
	\label{Figure:comp-sleep}
\end{figure*}

Fig.~\ref{Figure:comp-sleep} shows the hourly histogram of sleep durations in a day, of different elderly care center citizens. From these charts, we can infer different individuals sleeping patterns throughout the day. For instance, both charts show that individual 17 seems to be sleeping for a relatively longer period compared to other citizens, especially in the early hours. Similarly, individual 3 seems to be sleeping fairly long compared to others within the same duration. On the other hand, individual 0 seems to be significantly active during the early hours, as is visible from the reasonably low blue bars on the left side of the charts.

\begin{figure*}[h]
	\begin{center}
		\includegraphics[width=0.9\textwidth]{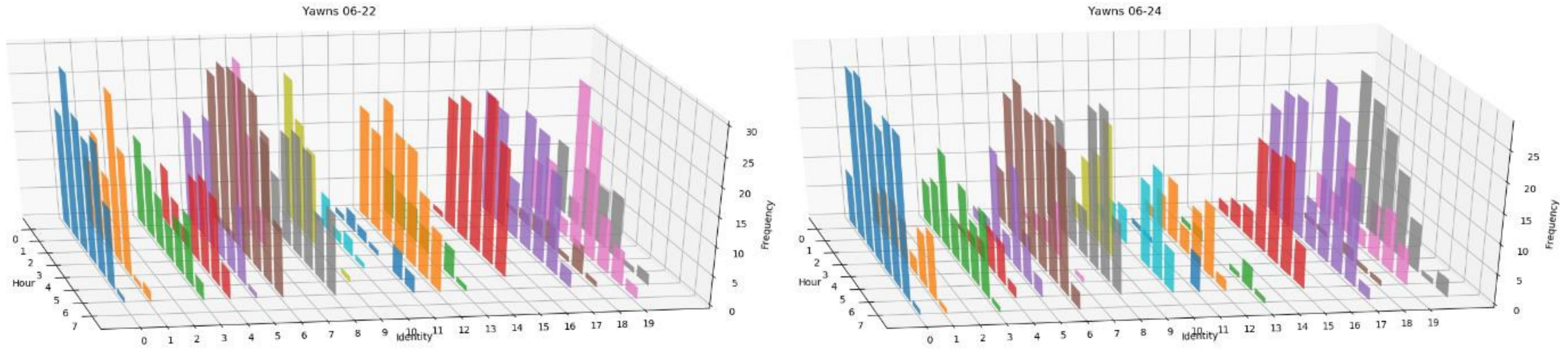}
	\end{center}
	\caption{Comparison of yawning time.}
	\label{Figure:comp-yawn}
\end{figure*}

The yawning frequency is shown in Fig.~\ref{Figure:comp-yawn}. Although yawn frequencies by itself may not give a lot of information, this information gives insight into the elderly citizens’ sleep during the periods. For instance, individual 17 had a fairly high yawn frequencies on both charts. This confirms observation from the previous figures: this individual indeed had slept for a longer duration compared to others. Contrastingly, individual 0 also experienced significantly higher yawns throughout the day compared to others, yet within the same period, sleep durations were dramatically lower compared to individual 17. We can then infer that this individual may have felt sleepiness, hence yawning, yet did not or could not fall asleep. These examples show that yawn information may either confirm or gives more insight into the sleeping pattern of the elderly care citizens.

These two analyses are important due to several reasons. Firstly, from these observations, caretakers may be able to infer that certain individuals may be lacking sleep during the night, hence allowing them to take appropriate actions against this, such as increasing their nightly hours. On the other hand, researchers have also shown that an inappropriate amount of sleep may affect memory adversely, which is crucial for patients afflicted with dementia. For citizens of an elderly care center, this is especially important as they are either at risk or have been afflicted with cognitive impairment. Thus, sleeping pattern throughout the day forms an incredibly important insight for the caretakers.

\begin{figure*}[h]
	\begin{center}
		\includegraphics[width=0.9\textwidth]{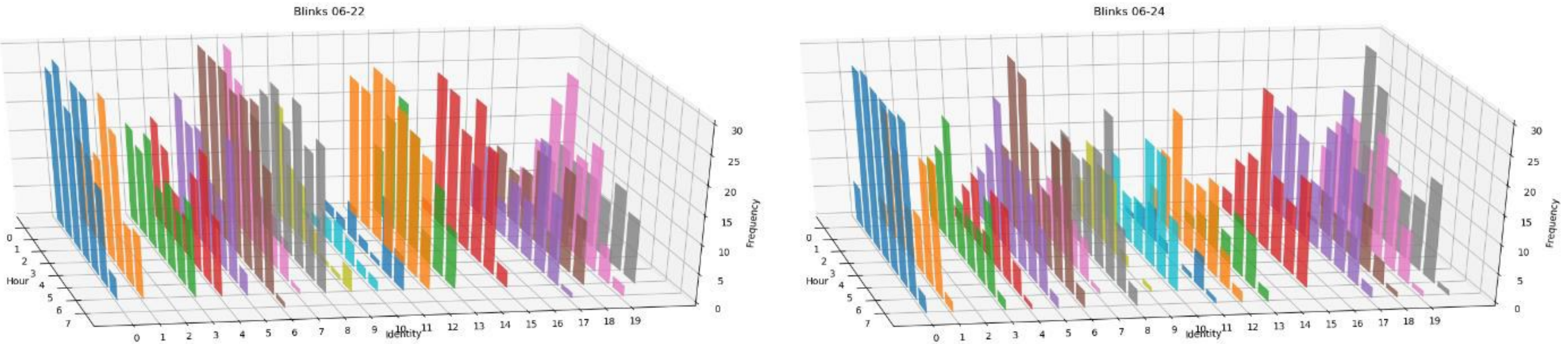}
	\end{center}
	\caption{Comparison of blinking time.}
	\label{Figure:comp-blink}
\end{figure*}

Fig.~\ref{Figure:comp-blink} show the hourly histogram of blink frequency in a day, of different elderly care center citizens. From this chart, it can be seen that individual 0 and individual 6 exhibit a relatively higher eye blink rate compared to other individuals across two days. However, overall, almost all the individuals show relatively active eye movements across the different data points, which is understandable since blinking is a natural and instinctive activity that humans possess.

Eyeblink rate (EBR), although seemingly trivial, has been shown to be related to mild cognitive impairment (MCI), as was alluded to earlier in the Literature Survey section. Since it was shown that an abnormally high EBR may be a marker for the transition to dementia, these observations are crucially important for the wellness of elders in the care center. Therefore, from these visualizations which show the eye blink rate of different individuals, experts in the field of mental health issues such as dementia may be able to take appropriate actions if accompanied by other markers.

All in all, the sleeping patterns and blinking patterns automatically obtained from our project will be very helpful to the caretakes of the elderly care center, since it allows them to learn more information regarding their citizens, which otherwise would have taken hours of manual observation. Although this information might not immediately cause an action to be taken, it shows signals which can then be followed up with closer and more reliable observation, done by experts in the field.

\subsection{Additional Analysis on Technical Challenges} 
%what need to be done/ new problem/technical enhancement for accurate and fully-automated solution
In this section, we will analyze more technical challenges in our system and possible technical enhancement for an accurate and fully automated solution in addition to our paper. These challenges include, but not limited to incorporating more fine-grained metrics, efficiency, generalization, and interpretability.

We output the temporal heatmap for caregivers to help them better understand the health condition. This is still one step from the fully-automated solution for healthcare as our system cannot directly output the score of health, attention level, and the alertness level. Defining the condition of health and attention requires much domain knowledge and we believe the current methods still fail to outperform the decision of professional caregivers. However, giving these score as a meta feature for some actions are feasible, such as giving the attention level when a senior citizen is watching TV. We can achieve the prediction via analysis on the temporal feature of the action. We are still working on it and plans to incorporate this module in next version.

Deep learning is much more effective compared to traditional image processing techniques in terms of inference. However, computational cost and facility in training is the main disadvantage. We train and evaluate the different models of our system on 2 NVIDIA TITAN RTX GPU. The computational environment is not applicable in everyone's home, makes it difficult to generalize the system to the houses of elderly individuals. Also, when we are processing the video, we run the detection algorithm frame by frame and aggregate the information with a sliding average. However, the information is highly redundant, which aggravates the burden to the computation process. Overall, we are also trying on a more effective model and we will strike a balance between efficiency and effectiveness. 

Our system shows promising effects on the data we collected. We utilize much information such as building a facial library for human recognition. We also give each elderly unique ID to better track the wellness condition. However, if we want to transfer our system to another application scenario (such as a household healthcare system), the performance will be limited because our model is highly specified. To the best of our knowledge, the research on the generalization of deep learning model is limited, which raises the concern that our model cannot perform well on another healthcare application setting.

Deep learning is a black box with powerful representation capacity. While DNNs can accurately classify and localize objects, yet still difficult to reason how it works in detail. The lack of Interpretability makes it vulnerable to adversarial attacks and brings the concern of fairness and privacy, especially for healthcare system. In our system, we use neural networks at almost all stages. When our system makes a decision such as encouraging higher exercise intensity, it is hard to explain how our model makes this decision, which is different from experienced caregivers who have abundant domain knowledge.

\end{document}